\documentclass[
aps,%
12pt,%
final,%
notitlepage,%
oneside,%
onecolumn,%
nobibnotes,%
nofootinbib,%
superscriptaddress,%
noshowpacs,%
centertags]%
{revtex4}

 \newcommand{\be}{\begin{equation}}
 \newcommand{\ee}{\end{equation}}
 \newcommand{\bea}{\begin{eqnarray}}
 \newcommand{\eea}{\end{eqnarray}}

\begin{document}
\selectlanguage{english}

\title{Flag-dipole and flagpole spinors fluid flows in Kerr spacetimes}

\author{\firstname{Rold\~{a}o}~\surname{da Rocha}}
\email{roldao.rocha@ufabc.edu.br}
\affiliation{%
CMCC, 
Universidade Federal do ABC, 09210-580, Santo Andr\'e - SP, Brazil.}
\affiliation{International School for Advanced Studies (SISSA), Via Bonomea 265, 34136 Trieste, Italy}
\author{\firstname{R.~T.}~\surname{Cavalcanti}}
\email{rogerio.cavalcanti@ufabc.edu.br}
 \affiliation{
CCNH, 
Universidade Federal do ABC,\\ 09210-580, Santo Andr\'e - SP, Brazil}
 \affiliation{
Dipartimento di Fisica e Astronomia, Universit\`a di Bologna, 
via Irnerio 46, 40126 Bologna, Italy}


\begin{abstract}
Flagpole and flag-dipole spinors are particular classes of spinor fields that has been recently used in different branches of theoretical physics. In this paper, we study the possibility and consequences of these spinor fields to induce an underlying fluid flow structure in the background of Kerr spacetimes. We show that flag-dipole spinor fields are solutions of the equations of motion in this context. To our knowledge, this is the second time that this class of spinor field  appears as a physical solution, the first one occurring as a solution of the Dirac equation in ESK gravities.
\end{abstract}

\maketitle

\section{Introduction}

There exists a robust spinor classification, based on the spinor bilinear covariants \cite{lou}, and built up over irreducible spin-1/2 representations of the Lorentz group. Within this classification, the first three classes are associated with   {{}{regular}} spinors. The fifth class encompasses Majorana and Elko spinors, while the sixth {{}{class, consisting of dipole spinors,}} covers Weyl spinors. {}Relevant to our purposes here, the fourth and fifth classes, according to \cite{lou}, comprise spinors supporting Penrose flags \cite{penrose}.  Indeed, the fifth class is  characterised by  flagpole  spinors -- presenting the electromagnetic momentum bivector as the flag and the current vector as the pole --  whilst the fourth class is determined by the flag-dipole spinors -- that have the electromagnetic momentum bivector as the flag and the current vector {\it and} the spin direction as two poles \cite{lou}. 
We can find in the literature a myriad of examples regarding the  classes of regular spinors, that encompass Dirac standard spinors, which are eigenspinors of the parity operator. In addition,  some singular spinors are comprised by the so-called classes five and six in Lounesto spinor fields classification. In particular, Majorana and Elko one-dimensional massive spinors are included in class five, whereas Weyl spinors are allocated in the sixth class of dipole spinors. Although we know physical examples of dynamical equations, providing these spinors, there are some unknown spinors in each one of the six classes. They encompass mass dimension one spinors and such novel possibilities can shed new light on our current knowledge in physics. The eigenspinors of the charge conjugation operator having dual helicity reveal unexpected new features that have been recently explored \cite{allu,Ahluwalia:2010zn,Ahluwalia:2008xi,Boehmer:2008rz,m1}.

Further than the already explored classes five and six of flagpoles and dipoles spinors, reside the flag-dipole spinors of class four. Although the scalar and pseudo-scalar bilinear covariants vanish in such class, such singular spinors are quite similar to regular spinors. From the dynamical point of view, it had not been a surprising fact that it could be generated as a solution of the Dirac in some particular situation. It was precisely what 
we found on 2013 to be the prototypical physical example of a flag-dipole spinor field \cite{fabri}, which was the less popular  type of spinors in Lounesto classification. Actually, solutions of the Dirac equation in Einstein-Sciama-Kibble gravity, that consists of a $f(R)$ theory of gravity in Einstein-Cartan spacetimes, satisfy the imperative constraints to be  a type-(4) flag-dipole singular spinor \cite{fabri,jul}. In spite of this single existing precedent that has been found two years ago, other type-(4) spinors are unknown, in particular these ones that arise as solutions of some equation of motion.
The program that has been paved across the last decade (check, for example, refs. \cite{Cavalcanti:2014wia,Villalobos:2015xca,daRocha:2011yr,Bernardini:2012sc} and references therein)  has the goal of determining the dynamics of all spinors  at each class in Lounesto  classification \cite{lou}.  {{}{Although we already know 
the general form of all kinds of spinors in Lounesto's classification \cite{Cavalcanti:2014wia}, the dynamics  for every  spinors lacks, still. Mass dimension one fermions 
\cite{Villalobos:2015xca} can occupy both classes four and five of spinor fields}}.

The goal of this paper has been motivated not only  by the prominent results on 4D Minkowski spacetime. In fact, 7D asymptotically flat black hole
backgrounds can provide a framework to spinors, whose  current
density defines a time-like Killing vector at the spatial
infinity. Flagpole spinors, typified by Elko spinors, have been already studied in black hole backgrounds, with the aim to study their Hawking radiation through tunnelling methods \cite{daRocha:2014dla}.  
Riemannian and pseudo-Riemannian 7D manifolds admit generalised classes of singular spinor fields  \cite{bonora,Bonora:2015ppa}.
 It is also
possible to see these spinors as
{soliton-like solutions} in a specific black hole background and to  identify
the current density components $J^\mu = \bar\psi\gamma^\mu\psi$ with the Killing vector field components at
the black hole horizon \cite{mei10}. This is inspired by the example of 
the Kerr and Myers-Perry 5d black holes. The latter constitutes an appropriate
background, where the current density interpolates between the time Killing vector
at the spatial infinity and the null Killing vector field on the black hole
event horizon \cite{mei10}. This current density can be seen as a spinor
fluid flow. In this paper we consider the extension of this construction
to Kerr spacetimes \cite{Bonora:2015ppa}. 
The only assumption consists of a spinor,  
whose spin density is coupled to the Riemann  tensor \cite{meifluid}. Therefore, an underlying fluid flow can be obtained, in particular, when flagpole and flag-dipole spinors are considered.

This work is organised as follows:  In the next section we revise the multivector structure underlying the theory of spinor fields, including the classification of spinors, accordingly. In section 3 we apply the constraints, that arise from the spinors classification,  to spinor fluid flows in Kerr spacetimes, hence extending the results of \cite{mei10,meifluid}. We finish the paper summarising and discussing further results. 

\section{Classifying spinors }

Throughout this paper, spinors take place in 4D Minkowski spacetime, as classical objects that carry a representation of the Lorentz group $\mathrm{Spin}^{e}_{1,3}\simeq\mathrm{SL}(2, \mathbb{C})$ in the space $\mathbb{C}^{\mathrm{4}}$. Indeed,  they are sections of the bundle $\mathbf{P}_{\mathrm{spin}_{\mathrm{1,3}}^{e}}(\textit{M})\times_{\rho}\mathbb{C}^{\mathrm{4}}$, where $\rho$ stands for the $\mathcal{D}^{\mathrm{(1/2, 0)}}\oplus\mathcal{D}^{\mathrm{(0, 1/2)}}$  representation of $\mathrm{SL}(\mathrm{2, \mathbb{C}})\simeq\mathrm{Spin}^{\textit{e}}_{1,3}$ in $\mathbb{C}^{\mathrm{4}}$ \cite{hidden}.
The set 
$
\{\gamma_{\mu}\}
$
 are the standard gamma matrices.
A multivector structure 
\begin{eqnarray*}
\mathrm{Z}=\sigma+\mathbf{J}+\textit{i}\mathbf{S}+\textit{i}\mathbf{K}\gamma_{\mathrm{0123}}
+\omega\gamma_{\mathrm{0123}},
\end{eqnarray*} where $\gamma_{\mathrm{0123}} = \gamma_0\gamma_1\gamma_2\gamma_3$, has each homogeneous part regarded to usual observables in Dirac theory, by
\begin{subequations}\begin{eqnarray}
\sigma&=&\bar\psi \psi,\\
\omega&=&-\bar\psi \gamma_{0123}\psi\\
{\mathbf{J}}=J_\mu{\bf e}^\mu&=&\bar\psi \gamma_{\mu}\psi {\bf e}^{\mu}\\
{\bf{K}}=K_\mu{\bf e}^\mu&=&\bar\psi \textit{i}\mathrm{\gamma_{0123}}\gamma_{\mu}\psi {\bf e}^{\mu}\\
{\mathbf{S}}=S_{\mu\nu}{\bf e}^\mu\wedge{\bf e}^\nu&=&\bar\psi {\it i}\gamma_{\mu\nu}\psi {\bf e}^{\mu}\wedge{\bf e}^{\nu}\end{eqnarray}
\end{subequations}\noindent that are interpreted for the electron, respectively, as the mass term, the pseudo-scalar, the current density, the spin direction, and the angular momentum distribution density.

Lounesto arranged all spinors in Minkowski spacetime in six different classes, according to the bilinear covariants for a given spinors. The bounds induced by the Fierz identities, introduced below, derive such six classes \cite{lou}: 
\begin{itemize}
\item[1)] $\sigma\neq0,\;\;\;\omega\neq0$\qquad\qquad\qquad\qquad\qquad4) $
\mathbf{K}\neq 0,\;\;\;\mathbf{S}\neq0$\;\;\; $(\sigma=\omega=0)$%
\label{Majorana 11}
\item[2)] $\sigma\neq0,\;\;\;
\omega = 0$\label{dirac1}\qquad\qquad\qquad\qquad\qquad5) $\mathbf{K}=0,\;\;\;
\mathbf{S}\neq0$\;\;\;$(\sigma=\omega=0)$
\label{tipo41}
\item[3)] $\sigma= 0, \;\;\;\omega \neq0$\label{dirac21}
\qquad\qquad\qquad\qquad\qquad\!6) $\mathbf{S}=0,
\;\;\; \mathbf{K} \neq 0$\;\;\;$(\sigma=\omega=0)$
\end{itemize}
\noindent  
  A spinor such that $\sigma\neq 0$ and/or $\omega\neq 0$ is said a regular spinor, whereas when $\sigma=0=\omega$, a spinor is said to be singular \cite{lou}. The  first three classes are composed by regular spinors and called regular type-1, 2 and 3 spinors. The fifth class, the flagpole, is represented by Majorana and Elko spinors \cite{rocha}, whereas the sixth class comprises Weyl spinors. The fourth class, the flag-dipole, has had its first physical example recently discovered \cite{fabri}. Some classes can be mapped into each other \cite{daRocha:2007pz}.

 There are some remarkable features associated to the $\mathrm{Z}$ multivector. First, the Fierz identities \cite{craw}
\begin{eqnarray}
&{\mathbf{J}}^2=\sigma^{2}+\omega^{2},\;\;\;\;\;\;\;\;-{\mathbf{K}}^{2}={\mathbf{J}}^{2},\;\;\;\;\;\;\;\;{\mathbf{J}}\cdot{\mathbf{K}}=0,\;\;\;\;\;\;\;\;{\mathbf{K}}\wedge{\mathbf{J}}=(\omega+\sigma\gamma_{\mathrm{0123}})\mathbf{S} \label{rfg}
\end{eqnarray} are valid for regular spinors, or whatever situation  that either $\bar\psi \psi\neq0$ or $\bar\psi \gamma_{0123}\psi\neq0$. For the case of singular spinors, Fierz identities are more general and substitute the standard ones (\ref{rfg}), that do not hold anymore. The generalised Fierz identities read:
\begin{eqnarray}
\mathrm{Z}^{2}/4\sigma=\mathrm{Z}, \;\;\;\; \quad4{J}^{\mu}\mathrm{Z}=\mathrm{Z}\gamma^{\mu}\mathrm{Z}, \;\;\;\;\quad 4{S}^{\mu\nu}\mathrm{Z}=i\mathrm{Z}\gamma^{\mu}\gamma^{\nu}\mathrm{Z}, \nonumber\\
-\mathrm{Z}\textit{i}\gamma^{\mu}\gamma^{\mathrm{0123}}\mathrm{Z}=4\mathrm{K}^{\mu}\mathrm{Z}, \;\;\;\;\quad  \mathrm{Z}\gamma^{\mathrm{0123}}\mathrm{Z}=-4\omega\mathrm{Z}.\label{Z}
\end{eqnarray} It is worth to mention that, in the case where the observables $\sigma, \mathbf{J}, \mathbf{S}, \mathbf{K},\omega$ satisfy the Fierz identities, the multivector $\mathrm{Z}$ is denominated a Fierz aggregate.  
The Z aggregate plays a prominent role in the framework here presented. Being observables, the bilinear covariants endow a physical interpretation for the multivector structure.

\section{Flag-dipole and flagpole spinor fluids}

A fluid can be essentially categorised  by a density, denoted by $\rho$, and a 4-velocity with components $u^\mu$ such that  $g_{\mu\nu}u^\mu u^\nu=-1$. When matter fields present angular momentum, as for instance a (Kerr) black hole, there is a dual description of the underlying fluid by  a  global current density. The vorticity is related to the exterior derivative of a fluid flow and the Riemann
curvature tensor as well, being coupled through 
the spin density \cite{meifluid}. In fact, this coupling is described by
\be d(\rho u)=\underbrace{i\bar\psi\gamma^{\mu\nu}\psi}_{S^{\mu\nu}} R_{\mu\nu
\rho\sigma} dx^\rho\wedge dx^\sigma\,.\label{rhou}\ee\noindent This choice is natural, in the sense that both
$d(\rho u)$ and $S^{\mu\nu}=i\bar\psi\gamma^{\mu\nu}\psi$ describe 
angular momenta, namely, the vorticity and the distribution of intrinsic angular momentum (or spin) density, respectively \cite{meifluid}. 
The spinor field $\psi$ is related to a time-like Killing vector $c_\psi{\bf J}$
at the spatial infinity, by
\be \lim_{r\rightarrow+\infty}\bar\psi\gamma^\mu\psi= -c_\psi J^\mu\,,\label{corrente}\ee
where $c_\psi >0$ is a normalisation constant. For
(\ref{rhou}) to hold,  the following
constraint on the spinor field  must be imposed \cite{meifluid}: 
\be  R_{\mu\nu\rho\sigma}\partial_\alpha (S^{\mu\nu})dx^\alpha\wedge dx^\rho\wedge dx^\sigma=0\,.
\label{psi.constr.2}\ee

Kerr black hole solutions can illustrate
what may be obtained from (\ref{rhou}), if the metric $ds^2=\eta_{ab} e^a_{~\mu}e^b_{~\nu} dx^\mu dx^\nu
=\eta_{ab}e^ae^b$ is regarded, as
\begin{eqnarray}
e^1&=&\sqrt{\frac{r^2+a^2\cos^2\theta}{\Delta}}\,dr\,,\nonumber\\
e^2&=&\sqrt{r^2+a^2\cos^2\theta}\;d\theta\,,\\ e^3&=&\frac{(r^2+a^2)\sin\theta}{\sqrt{r^2+a^2\cos^2\theta}}\,\left(d\varphi-\frac{a}{r^2+a^2}
dt\right)\,,\\
 e^4&=&\sqrt{\frac{\Delta}{r^2+a^2\cos^2\theta}}\left(dt-a\sin^2\theta\,d\phi\right),\quad \text{where}\quad\Delta=r^2-2Mr+a^2\,.\label{metric}\end{eqnarray}
The gamma matrices, in this context, read
\be\gamma^a=-i\sigma_a\otimes\sigma_2\,,\quad
\gamma^4=i 1_2\otimes \sigma_3\,,\;\;\;\;a=1,2,3\,,\label{gamma.matrices}\ee
and the spinor field can be expressed as
\bea\psi=\left(\psi_{1a}+i\psi_{1b}, 
\psi_{2a}+i\psi_{2b}, \psi_{3a}+i\psi_{3b},  \psi_{4a}
+i\psi_{4b}\right)^\intercal\,,\label{ansatz.psi}\eea
where all real functions in (\ref{ansatz.psi}) are $(r,\theta)$-dependent. The metric (\ref{metric}) yields the components of the current density: 
\bea\bar\psi\gamma^r\psi&=&-2\sqrt{\frac{r^2+a^2\cos^2\theta}{\Delta}}\,z_1\,,\nonumber\\ \bar\psi
\gamma^\phi \psi
 &=&-\frac{1}{\sqrt{r^2+a^2\cos^2\theta}}\Big(\frac{2z_3}{\sin\theta}+\frac{az_4}{\sqrt{{\Delta}}} \Big)\,,\nonumber\\
\bar\psi\gamma^\theta\psi&=&-\frac{2}{\sqrt{r^2+a^2\cos^2\theta}}\,,\nonumber\\
\bar\psi\gamma^t\psi&=&-\frac{r^2}{\sqrt{r^2+a^2\cos^2\theta}}\Big(\frac{2a\sin\theta z_3}{r^2+a^2}+\frac{z_4}{\sqrt{{\Delta}}} \Big)\,,\eea
with
\bea z_1&=&(\psi_{2b}\psi_{3b}+\psi_{2a}\psi_{3a}  +\psi_{1b}\psi_{4b}+\psi_{1a}
\psi_{4a})\,,\nonumber\\
z_2&=&(\psi_{2b}\psi_{3a} -\psi_{2a}\psi_{3b} \psi_{1a}\psi_{4b}-\psi_{1b}\psi_{4a}
)\,,\nonumber\\
z_3&=&(\psi_{1a}\psi_{3a} -\psi_{2a}\psi_{4a}+\psi_{1b}\psi_{3b}
-\psi_{2b}\psi_{4b})\,,\nonumber\\
z_4&=&(\psi_{1b}^2 +\psi_{1a}^2+\psi_{2b}^2 +\psi_{2a}^2
+\psi_{3b}^2 +\psi_{3a}^2+\psi_{4b}^2+\psi_{4a}^2)\,.\eea
For the Kerr metric (\ref{metric}), the time-like Killing vector
is $J^\mu\partial_\mu=\partial_t$. Now, eq.  (\ref{corrente}) can be satisfied with
\be z_1=z_2=z_3=0\,,\qquad\quad z_4=c_\psi \frac{\sqrt{\Delta}(r^2+a^2)}{(r^2+a^2\cos^2\theta)^{3/2}}\,.
\label{tresmeios}\ee
The first three equations are easily solved by
\be\psi_{3b}=-\frac{\psi_{3a}}{\psi_{1b}}\psi_{1a}\,,\quad
\psi_{4a}=\frac{\psi_{3a}}{\psi_{1b}}\psi_{2b}\,,\quad
\psi_{4b}=-\frac{\psi_{3a}}{\psi_{1b}}\psi_{2a}\,.
\label{spinor1}\ee

By imposing that the spinor is singular, it implies that
\be
\|\psi_3\|^2 = \psi_{4a}\left(\frac{\psi_{2a}}{\psi_{2b}}\psi_{4b}-\psi_{4a}\right)\ee
where $\psi_\mu = \psi_{\mu a} + i\psi_{\mu b}$.
The conditions that singular spinors are either flag-dipole or  flagpole spinors are: $\|\psi_3\|^2 \neq \|\psi_2\|^2$ and $\|\psi_3\|^2 = \|\psi_2\|^2$, respectively. In the first case, we achieve a new physical example of a flag-dipole spinor, regarding a spinor in a Kerr black hole background. Moreover, the spinor provides by Eq.(\ref{corrente}) a current density $\bar\psi\gamma^\mu\psi$ that approaches a time-like Killing spinor at the spatial infinity.

Eq. (\ref{rhou}) implies that 
\be d(\rho u)=f_{rh}dr\wedge d\theta +f_{pt}d\varphi\wedge dt
+f_{hp}d\theta\wedge d\varphi +f_{rp}dr\wedge d\varphi
+f_{ht}d\theta\wedge dt +f_{rt}dr\wedge dt\,,\label{du.1}\ee
with
\bea f_{rh}&=&\frac{r^2+a^2\cos^2\theta}{\sqrt\Delta}M\left({12ar_1\cos\theta}\kappa_1
+{2r\,r_3}\kappa_2\right)\,,\nonumber\\
f_{pt}&=&\frac{2M\sin\theta\sqrt{\Delta}}{(r^2+a^2\cos^2\theta)^{3}}\left({2rr_3\sin\theta}\kappa_1
-{3ar_1\cos\theta}\kappa_2\right)\,,\nonumber\\
f_{hp}&=&\frac{4M\sin\theta}{(r^2+a^2\cos^2\theta)^{3}}\left({arr_3\sin\theta\sqrt{\Delta}}\kappa_3
-{2r(r^2+a^2)r_3}{}\kappa_4\right.\nonumber\\
&&-\left.{6a(r^2+a^2)r_1\cos\theta}\kappa_5
-{3a^2r_1\cos\theta\sqrt{\Delta}}\kappa_6\right)\,,\nonumber\\
f_{rp}&=&\frac{4M\sin\theta}{\sqrt\Delta(r^2+a^2\cos^2\theta)^{3}}\left({3a(r^2+a^2)r_1\cos\theta}\kappa_3-{6a^2r_1\cos\theta\sin\theta\sqrt{\Delta}}\kappa_4\right.\\
&&\left.+{2arr_3\sin\theta}\kappa_5
+{r(r^2+a^2)r_3}{\sqrt{\Delta}}\kappa_6\right)\,,\nonumber\\
f_{ht}&=&\frac{4M}{(r^2+a^2\cos^2\theta)^{3}}\left(-{rr_3\sqrt{\Delta}}\kappa_3
+{2arr_3\sin\theta}\kappa_4
+{6a^2r_1\cos\theta\sin\theta}\kappa_5+{3ar_1\cos\theta\sqrt{\Delta}}\kappa_6\right)\,,\nonumber\\ \nonumber
f_{rt}&=&\frac{4M}{\sqrt\Delta(r^2+a^2\cos^2\theta)^{3}}\left(-\frac{3a^2r_1\cos\theta\sin\theta}{\sqrt{\Delta}} \kappa_3
+{6ar_1\cos\theta}{}\kappa_4 -{2rr_3}\kappa_5
-\frac{arr_3\sin\theta}{\sqrt{\Delta}}\kappa_6\right)\,,\label{def.vv}\eea
where $r_1=r^2-\frac13a^2\cos^2\theta$, $r_3=r^2
-3a^2\cos^2\theta$ and
\bea \kappa_1&=&\psi_{1b}\psi_{3a}-\psi_{1a}\psi_{3b}
-\psi_{2b}\psi_{4a}+\psi_{2a}\psi_{4b}\,,\nonumber\\
\kappa_2&=&\psi_{1a}^2+\psi_{1b}^2-\psi_{2a}^2-\psi_{2b}^2
-\psi_{3a}^2-\psi_{3b}^2+\psi_{4a}^2+\psi_{4b}^2\,,\nonumber\\
\kappa_3&=&\psi_{2a}\psi_{3a}+\psi_{2b}\psi_{3b}
-\psi_{1a}\psi_{4a}-\psi_{1b}\psi_{4b}\,,\nonumber\\
\kappa_4&=&\psi_{1a}\psi_{2a}+\psi_{1b}\psi_{2b}
-\psi_{3a}\psi_{4a}-\psi_{3b}\psi_{4b}\,,\nonumber\\
\kappa_5&=&\psi_{2b}\psi_{3a}-\psi_{2a}\psi_{3b}
+\psi_{1b}\psi_{4a}-\psi_{1a}\psi_{4b}\,,\nonumber\\
\kappa_6&=&\psi_{1b}\psi_{2a}-\psi_{1a}\psi_{2b}
-\psi_{3b}\psi_{4a}+\psi_{3a}\psi_{4b}\,.\label{cond1}\eea
The author in \cite{meifluid} argues that $\rho u$ has a physically significance if the constraint 
$f_{pt}=0$ in (\ref{du.1}) is imposed. Hence 
\be \psi_{1a}^2+\psi_{1b}^2=\psi_{2a}^2+\psi_{2b}^2\,.\label{quadrado}\ee
Eqs. (\ref{tresmeios}) and (\ref{quadrado}) can be
solved by
\be\psi_{1a}=\sqrt{\psi_{2a}^2+\psi_{2b}^2-\psi_{1b}^2}\;,\qquad
\psi^2_{3a}=\psi^2_{1b}{\frac{c_\psi }{2 \left(\psi_{1a}^2+\psi_{1b}^2\right)^2}
\frac{(r^2+a^2\cos^2\theta)^{3/2}}{\sqrt{\Delta}(r^2+a^2)}}\;.\label{comps1}\ee
Now $\psi_{1b}$,
$\psi_{2b}$ and $ \psi_{1a}^2+\psi_{1b}^2$ remain, since  degrees of freedom arise from $d^2(\rho u)=0$ \cite{meifluid}.

By taking into account the 
interesting
case is the limit of a lower rotating black hole  
$a\rightarrow0$, Ref. \cite{meifluid} adopts the expansion 
\bea\psi_{1b}&=&\psi_{10}+\psi_{11}a+{\mathcal O}(a^2),\nonumber\\
\psi_{2b}&=&\psi_{20}+\psi_{21}a+{\mathcal O}(a^2)\,,\eea
namely 
\bea\psi_{20}&=&\psi_{10}\;,\nonumber\\
\psi_{21}&=&\psi_{11}\,,\nonumber\\
\psi_{22}&=&\psi_{12}\!-\!\frac{r^2
u_{t2}^\prime(\theta)}{4Mf} \sqrt{c_\psi 
\sqrt{f}\!-\!4\psi_{10}^2} \!-\!\sin(2\theta)\sqrt{c_\psi 
\sqrt{f}\!-\!4\psi_{10}^2}\\ \nonumber
&&\!\!\!\!\!\!\!\!\!\!\!\!\!\!\!\!\!\!\!\!\!\!\!\left(\frac{2(1260M^4\!-\!180M^3r\!-\!213M^2r^2 \!-\!142r^3(M\!-\!r)}{23625M^4 r\sqrt{\Delta}}\!+\!\frac{r(9M^2\!+\!12Mr\!+\!50 r^2)\!+\!24r^3\ln(1 +\sqrt{f})}{675 M^4
\sqrt{f \Delta}}\right)\,,\label{result.psi}\eea
where $f=1\!-\!\frac{2M}r$. A similar technique was employed in \cite{Bonora:2012eb}.

In the limit $a\rightarrow0$, the black hole horizon is located at
$r_0\approx2M-\frac{a^2}{2M}$. The result in (\ref{result.psi})
diverges at $r=2M$, however the framework is important in the  region  outside the black hole, where $r>2M$. Hence this divergence does not take place for the framework here presented. 

A fluid flow can be attained in such a framework. In fact Eq. (\ref{rhou}) implies that 
\bea\rho&=&\frac{4c_\psi }{15m}f\Big(1+\frac{3M}r\Big) +{\mathcal O}(a^2)\,,\nonumber\\
u^t&=&\frac1{\sqrt{f}}+{\mathcal O}(a^2)\,,\nonumber\\\quad u^r&=&u^\theta =0\,,\nonumber\\
u^\phi&=&\frac{a}{r^2 \sqrt{f}} \Big(1-\frac{1}{\sqrt{f}}\left(1+\frac{3M}r\right)^{-1}
\Big) +{\mathcal O}(a^2) \,.\label{fluid}\eea
Prominent features arise from this prescription.  In fact, since $\lim_{r\rightarrow +\infty} u=\partial_t$, there is a static fluid description at this limit. 
One also has $u^\phi\sim a$, which means that there
is no spatial flow underlying the Schwarzschild black hole. 
Moreover,  the fluid density vanishes exactly on the black hole 
horizon.  Now, if $\rho\rightarrow0$ at some point in
the {} spacetime, it does not mean that the density of the
fluid really vanishes. The limits  $\lim_{r\rightarrow
+\infty}u^\phi\sim r^{-2}$ and $\lim_{f\rightarrow
0}u^\phi=\infty$ also hold in this situation. Similarly, the time component of the velocity $u^t$ further diverges
at $f=0$. The divergence does not matter  outside the black hole horizon. Finally, $\lim_{r\rightarrow+\infty}\rho=4c_\psi /15M$, what implies that the density of
the fluid goes asymptotically the Minkowski spacetime \cite{meifluid}. 

The current density components $J^\mu=\bar\psi\gamma^\mu\psi$ is led by  (\ref{spinor1}) and (\ref{comps1}) to
\be {\bf J} = J_\mu {\bf e}^\mu =-c_\psi (f_a\partial_\phi+\partial_t)\,.\ee
The vector $c_\psi ^{-1}{\bf J}$ interpolates between the time-like
Killing vector at the spatial infinity and the null Killing vector
on the black hole horizon \cite{meifluid}. Furthermore,   by identifying the current of probability to the fluid as 
\be J^\mu =-\rho u^\mu\,,\label{corrfluid}\ee
then
\bea\rho&=&c_\psi \frac{\sqrt{{\Delta}(r^2+a^2\cos^2\theta)}}{r^2+a^2}\,,\nonumber\\
u&=&\frac{c_\psi }{\rho}\left(\frac{a}{a^2+r^2}\partial_\phi+\partial_t\right)\,. \label{fluid2}\eea
Hence, flagpole and flag-dipole spinors can generate a fluid 
(\ref{fluid2}) in the vicinity of a black hole. The spinor current density can be further identified 
to a Killing vector at the spatial infinity.

\section{Concluding remarks}

The underlying spinor fluid behaviour was studied considering  Kerr backgrounds. This fluid is static and uniform when $r\to\infty$,  nevertheless it presents vorticity when $r\sim 0$. Ref. \cite{meifluid}  considers stationary spacetimes, which have a  time-like Killing vector attached to them.
We already established the correspondence between the current density and the Killing vector 
at the spatial infinity in stationary axisymmetric black holes in 7D. Such Killing vector fields were shown to interpolate between the null Killing vector on the horizon and the time-like Killing vector at $r\to\infty$ \cite{wei,mei10}. 

\section*{Acknowledgments}
 RdR is grateful to CNPq grants No. 473326/2013-2, No. 451682/2015-7 and No. 303027/2012-6, to FAPESP Grant No. 2015/10270-0, to INFN project ``Classification of Spinors'', and to SISSA, for the hospitality. RTC is grateful to CAPES, PDSE and to Prof. R. Casadio for useful discussions and for the hospitality.

\end{document}